\documentclass[twoside,psfig]{article}
\usepackage{graphicx}
\catcode`\@=11 \long\def\@makefntext#1{ \protect\noindent \hbox to
3.2pt {\hskip-.9pt
$^{{\eightrm\@thefnmark}}$\hfil}#1\hfill}       

\def\@makefnmark{\hbox to 0pt{$^{\@thefnmark}$\hss}}    

\def\ps@myheadings{\let\@mkboth\@gobbletwo
\def\@oddhead{\hbox{}
\rightmark\hfil\eightrm\thepage}
\def\@oddfoot{}\def\@evenhead{\eightrm\thepage\hfil
\leftmark\hbox{}}\def\@evenfoot{}
\def\sectionmark##1{}\def\subsectionmark##1{}}

\oddsidemargin=\evensidemargin 
\newcounter{sectionc}\newcounter{subsectionc}\newcounter{subsubsectionc}
\renewcommand{\section}[1] {\vspace{12pt}\addtocounter{sectionc}{1}
\setcounter{subsectionc}{0}\setcounter{subsubsectionc}{0}\noindent
    {\tenbf\thesectionc. #1}\par\vspace{5pt}}
\renewcommand{\subsection}[1] {\vspace{12pt}\addtocounter{subsectionc}{1}
    \setcounter{subsubsectionc}{0}\noindent
    {\bf\thesectionc.\thesubsectionc. {\kern1pt \bfit #1}}\par\vspace{5pt}}
\renewcommand{\subsubsection}[1] {\vspace{12pt}\addtocounter{subsubsectionc}{1}
    \noindent{\tenrm\thesectionc.\thesubsectionc.\thesubsubsectionc.
    {\kern1pt \tenit #1}}\par\vspace{5pt}}
\newcommand{\nonumsection}[1] {\vspace{12pt}\noindent{\tenbf #1}
    \par\vspace{5pt}}

\newcounter{appendixc}
\newcounter{subappendixc}[appendixc]
\newcounter{subsubappendixc}[subappendixc]
\renewcommand{\thesubappendixc}{\Alph{appendixc}.\arabic{subappendixc}}
\renewcommand{\thesubsubappendixc}
    {\Alph{appendixc}.\arabic{subappendixc}.\arabic{subsubappendixc}}

\renewcommand{\appendix}[1] {\vspace{12pt}
        \refstepcounter{appendixc}
        \setcounter{figure}{0}
        \setcounter{table}{0}
        \setcounter{lemma}{0}
        \setcounter{theorem}{0}
        \setcounter{corollary}{0}
        \setcounter{definition}{0}
        \setcounter{equation}{0}
        \renewcommand{\thefigure}{\Alph{appendixc}.\arabic{figure}}
        \renewcommand{\thetable}{\Alph{appendixc}.\arabic{table}}
        \renewcommand{\theappendixc}{\Alph{appendixc}}
        \renewcommand{\thelemma}{\Alph{appendixc}.\arabic{lemma}}
        \renewcommand{\thetheorem}{\Alph{appendixc}.\arabic{theorem}}
        \renewcommand{\thedefinition}{\Alph{appendixc}.\arabic{definition}}
        \renewcommand{\thecorollary}{\Alph{appendixc}.\arabic{corollary}}
        \noindent{\tenbf Appendix \theappendixc #1}\par\vspace{5pt}}
\newcommand{\subappendix}[1] {\vspace{12pt}
        \refstepcounter{subappendixc}
        \noindent{\bf Appendix \thesubappendixc. {\kern1pt \bfit #1}}
    \par\vspace{5pt}}
\newcommand{\subsubappendix}[1] {\vspace{12pt}
        \refstepcounter{subsubappendixc}
        \noindent{\rm Appendix \thesubsubappendixc. {\kern1pt \tenit #1}}
    \par\vspace{5pt}}

\newcommand{\textlineskip}{\baselineskip=13pt}
\newcommand{\smalllineskip}{\baselineskip=10pt}

\def\eightcirc{
\begin{picture}(0,0)
\put(4.4,1.8){\circle{6.5}}
\end{picture}}
\def\eightcopyright{\eightcirc\kern2.7pt\hbox{\eightrm c}}

\newcommand{\copyrightheading}[1]
    {\vspace*{-2.5cm}\smalllineskip{\flushleft
    {\footnotesize International Journal of Modern Physics C #1}\\
    {\footnotesize $\eightcopyright$\, World Scientific Publishing
     Company}\\
     }}

\newcommand{\publisher}[2]{{\begin{center}\footnotesize\smalllineskip
    Received #1\\
    Revised #2
    \end{center}
    }}

\def\abstracts#1#2#3{{
    \centering{\begin{minipage}{4.5in}\footnotesize\baselineskip=10pt
    \parindent=0pt #1\par
    \parindent=15pt #2\par
    \parindent=15pt #3
    \end{minipage}}\par}}

\def\keywords#1{{
    \centering{\begin{minipage}{4.5in}\footnotesize\baselineskip=10pt
    {\footnotesize\it Keywords}\/: #1
    \end{minipage}}\par}}

\newcommand{\bibit}{\nineit}
\newcommand{\bibbf}{\ninebf}
\renewenvironment{thebibliography}[1]
        {\frenchspacing
     \ninerm\baselineskip=11pt
         \begin{list}{\arabic{enumi}.}
        {\usecounter{enumi}\setlength{\parsep}{0pt}
     \setlength{\leftmargin 12.7pt}{\rightmargin 0pt} 
         \setlength{\itemsep}{0pt} \settowidth
    {\labelwidth}{#1.}\sloppy}}{\end{list}}

\newcounter{itemlistc}
\newcounter{romanlistc}
\newcounter{alphlistc}
\newcounter{arabiclistc}

\newcommand{\fcaption}[1]{
        \refstepcounter{figure}
        \setbox\@tempboxa = \hbox{\footnotesize Fig.~\thefigure. #1}
        \ifdim \wd\@tempboxa > 5in
           {\begin{center}
        \parbox{5in}{\footnotesize\smalllineskip Fig.~\thefigure. #1}
            \end{center}}
        \else
             {\begin{center}
             {\footnotesize Fig.~\thefigure. #1}
              \end{center}}
        \fi}

\newcommand{\tcaption}[1]{
        \refstepcounter{table}
        \setbox\@tempboxa = \hbox{\footnotesize Table~\thetable. #1}
        \ifdim \wd\@tempboxa > 5in
           {\begin{center}
        \parbox{5in}{\footnotesize\smalllineskip Table~\thetable. #1}
            \end{center}}
        \else
             {\begin{center}
             {\footnotesize Table~\thetable. #1}
              \end{center}}
        \fi}

\def\@citex[#1]#2{\if@filesw\immediate\write\@auxout
    {\string\citation{#2}}\fi
\def\@citea{}\@cite{\@for\@citeb:=#2\do
    {\@citea\def\@citea{,}\@ifundefined
    {b@\@citeb}{{\bf ?}\@warning
    {Citation `\@citeb' on page \thepage \space undefined}}
    {\csname b@\@citeb\endcsname}}}{#1}}

\newif\if@cghi
\def\cite{\@cghitrue\@ifnextchar [{\@tempswatrue
    \@citex}{\@tempswafalse\@citex[]}}
\def\citelow{\@cghifalse\@ifnextchar [{\@tempswatrue
    \@citex}{\@tempswafalse\@citex[]}}
\def\@cite#1#2{{$\null^{#1}$\if@tempswa\typeout
    {IJCGA warning: optional citation argument
    ignored: `#2'} \fi}}

\def\pmb#1{\setbox0=\hbox{#1}
    \kern-.025em\copy0\kern-\wd0
    \kern.05em\copy0\kern-\wd0
    \kern-.025em\raise.0433em\box0}

\def\fnt#1#2{\footnotetext{\kern-.3em
    {$^{\mbox{\scriptsize #1}}$}{#2}}}

\def\ps@myheadings{%
    \let\@oddfoot\@empty\let\@evenfoot\@empty
    \def\@evenhead{\slshape\leftmark\hfil}
    \def\@oddhead{\hfil{\slshape\rightmark}}
    \let\@mkboth\@gobbletwo
    \let\sectionmark\@gobble
    \let\subsectionmark\@gobble
    }
\font\tenrm=cmr10 \font\tenit=cmti10 \font\tenbf=cmbx10
\font\bfit=cmbxti10 at 10pt \font\ninerm=cmr9 \font\nineit=cmti9
\font\ninebf=cmbx9 \font\eightrm=cmr8

\def\qed{\hbox{${\vcenter{\vbox{            
   \hrule height 0.4pt\hbox{\vrule width 0.4pt height 6pt
   \kern5pt\vrule width 0.4pt}\hrule height 0.4pt}}}$}}

\def\bsc{{\sc a\kern-6.4pt\sc a\kern-6.4pt\sc a}}   
\def\bflatex{\bf L\kern-.30em\raise.3ex\hbox{\bsc}\kern-.14em
T\kern-.1667em\lower.7ex\hbox{E}\kern-.125em X}

\begin{document}
\setlength{\textheight}{7.7truein}  

\thispagestyle{empty}

\markboth{\protect{\footnotesize\it Mingfeng He et
al.}}{\protect{\footnotesize\it Evolution of population with
sexual and asexual reproduction in changing environment}}

\normalsize\textlineskip

\setcounter{page}{1}

\copyrightheading{}         

\vspace*{0.88truein}

\centerline{\bf Evolution of population with sexual and asexual
reproduction} \vspace*{0.035truein} \centerline{\bf in changing
environment } \vspace*{0.37truein} \centerline{\footnotesize
Mingfeng
 He} \baselineskip=12pt \centerline{\footnotesize\it
Department of Applied Mathematics, Dalian University of
Technology,}\centerline{\footnotesize\it
\centerline{\footnotesize\it   Institute of undergraduate
innovation, Dalian University of Technology,}} \baselineskip=10pt
\centerline{\footnotesize\it Dalian 116024, China}

\vspace*{0.37truein} \centerline{\footnotesize Hongbo Ruan}
\baselineskip=12pt \centerline{\footnotesize\it Department of
Civil Engineering, Dalian University of
Technology,}\centerline{\footnotesize\it
\centerline{\footnotesize\it Institute of undergraduate
innovation, Dalian University of Technology,}} \baselineskip=10pt
\centerline{\footnotesize\it Dalian 116024, China}
\centerline{\footnotesize\it E-mail: ruan\_123@tom.com}

\vspace*{0.37truein} \centerline{\footnotesize Changliang Yu}
\baselineskip=12pt \centerline{\footnotesize\it Department of
Automation, Dalian University of
Technology,}\centerline{\footnotesize\it
\centerline{\footnotesize\it Institute of undergraduate
innovation, Dalian University of Technology,}} \baselineskip=10pt
\centerline{\footnotesize\it Dalian 116024, China}

 \vspace*{0.37truein} \centerline{\footnotesize Lei
Yao} \baselineskip=12pt \centerline{\footnotesize\it Department of
Automation, Dalian University of Technology,} \baselineskip=10pt
\centerline{\footnotesize\it Dalian 116024, China}

\vspace*{0.225truein} \publisher{(received date)}{(revised date)}

\vspace*{0.25truein} \abstracts{Using a lattice model based on
Monte Carlo simulations, we study the role of the reproduction
pattern on the fate of an evolving population. Each individual is
under the selection pressure from the environment and random
mutations. The habitat ("climate") is changing periodically.
Evolutions of populations following two reproduction patterns are
compared, asexual and sexual. We show, via Monte Carlo
simulations, that sexual reproduction by keeping more diversified
populations gives them better chances to adapt themselves to the
changing environment. However, in order to obtain a greater chance
to mate, the birth rate should be high. In the case of low birth
rate and high mutation probability there is a preference for the
asexual reproduction.}{}{}

\vspace*{5pt} \keywords{Evolution; Sexual; Asexual; Aging.}


\vspace*{1pt}\textlineskip  
\section{Introduction}     
\vspace*{-0.5pt} \noindent Most of today's higher species prefer
sexual over asexual reproduction, and even asexual species often
employ some form of genetic recombination. There are many theories
but no consensus$^{1}$  why sex is preferred. The traditional
explanation for sex is that it accelerates adaptation by allowing
two or more beneficial mutations that have appeared in different
individuals to recombine within the same individual $^{2}$. Sex
produces variation and thereby promotes evolutionary adaptation.
But how does sex achieve this effect, and under what circumstances
is it worthwhile?

 In the Redfield model, computer simulations have given clear
  advantages of one or the other way, depending on the parameters.
  A genomic bit-string model without ageing $^{3,4}$ also justified sexual reproduction;
   in one case the asexual way of life even died out.
   Computer simulations based on the Penna model $^{5,6,9}$
   have shown that sexual reproduction leads to populations with
   larger variety of genotypes, although the survival rates for
   asexual reproduction could be the same $^{9}$. In particular Sa
    Martins and Moss de Oliveira $^{7}$ have shown that the sexual
     reproduction gives a better survival chances for a population
      in the case of a natural disaster.

Most of those models assumed that the environment is constant and
the mutations are detrimental, as is customary in ageing theories
$^{8}$. However, a question which is often asked by biologists,
but cannot be tested in nature is, what is the role of the changes
in the environment and of the mutations in the biological
evolution. Both are considered to be important factors in
determining the fate of an evolving population. To study the role
of the reproduction pattern in the fate of an evolving population,
it is necessary to investigate the role played by those two
factors. The present paper introduces the following assumptions.
In the constantly changing environment, we assume that the
mutations are blind. The mutations which are helpful to make the
population more adaptive to the environment are called beneficial
mutations, and those which are negative for the population to
adjust to the environment are called deleterious mutations. Below,
we present a Monte Carlo study basing on the above assumptions.

The present paper ignores the intermediate reproduction forms
("meiotic parthenogenesis" and "hermaphroditism", which are less
wide spread), and compares sexual (now abbreviated by SR) with
asexual reproduction (abbreviated by AR), assuming genetic
recombination only in the sexual case. We want to investigate the
problem of possible advantages of either of the two above
mentioned reproduction mechanisms.

The paper is organized as follows. In Sec. 2, we define our model.
In Sec. 3, we present our findings from the simulations. Finally
in Sec. 4, we state our conclusions.

\section{The  model }
\noindent Consider a square lattice $L \times L$ with periodic
boundary conditions, where initially $N(t=0)$ individuals are
located. Each of them is characterized by its location ($k$) at a
lattice cell, genotype ($g_k$), phenotype ($h_k$) and age ($A_k$).
On each lattice cell there might be at most one individual. The
space (habitat) as well as time is discrete.

The asexual version (haploid organisms) of the model considers a
single string of 32 sites (loci) associated to each individual of
a population, representing its genotype. For the sexual case
(diploid organisms), each individual has two such strings
(chromosomes) read in parallel. Each locus can be in two states,
called alleles, and denoted here as 0 and 1. Thus a genotype is a
single or double string of zeros and ones. They are defined at
birth, kept unchanged during the individual lifetime.

From a genotype a phenotype is constructed, as a single string of
 the same length, according to the following rule. For haploid
 organisms the phenotype is identical with the genotype.
  The diploid phenotype is constructed as follows - All the loci are sorted
  as recessive and dominant loci, each of which takes up 50\% of the total loci.
  On the dominant loci, if a "1" appears (dominant homozygote -- (1,1)
  and heterozygote-- (0,1) or (1,0)), a 1 is put at the corresponding place
  of the phenotype, otherwise a 0. While on the recessive loci, if a "0" appears
  (recessive homozygote--(0,0) and heterozygote-- (0,1) or (1,0)), a 0 is put at
  the corresponding place of the phenotype, otherwise a 1. We interpret the bit-string
  not only the phenotype but also 32 intervals ("years") in the life of an individual.
  At a certain age we see only the loci corresponding to this and earlier ages.
  Age $A_k$ is augmented for each individual after completion of one Monte Carlo step (MCS).

The environment is represented in the model by a certain "ideal"
phenotype, $H$, which ensures the best adaptation for an
individual to the existing conditions. Such "ideal" phenotype has
been already introduced in the literature $^{10, 11,12}$. The
fitness of an individual to the environment is measured as the
agreement of the individual's phenotype with the "ideal"
phenotype. Thus, there are no absolutely good or bad genes in our
model. Only when a locus on the phenotype is identical with the
corresponding locus on the "ideal" phenotype, it is considered to
be a good one and vice versa. The environment is constantly
changing. Therefore the "good" or "bad" for genes are also
temporal.

Here we consider random mutations affecting genotypes of the
offspring. On each locus of the genotype, a mutation may take
place with probability $M$.

The mutation occurs at a randomly chosen bit of the genotype of
the offspring, with the probability $M$.

Here we use separate computer simulations for sexual and asexual
reproduction. The process starts with an initial population,
$N(0)$, having random genotypes $g_k$, chosen from a uniform
distribution. The dynamical rule is given by the following steps.

(1). An individual $k$ is randomly chosen. In the case of SR, only
females are selected.

 (2). One search is done for an empty cell in the nearest neighborhood.
 If not found, the individual stays in place. Otherwise the individual moves to the new position.

(3). In the case of SR, if the age of the female is between $A_m$
and $A_d$, a male with age larger than $A_m$ is chosen in the
Moore neighborhood of the female. If it is found, the two give
birth to $q$ offspring, each of which is randomly either male or
female. For each of the offspring an independent search is made
for a free cell in the Moore neighborhood of the female. If found,
the progeny is put there. If the cell is already occupied, nothing
happens and we search for a place for the next offspring.

In AR no mate is needed and if the age of the individual is
between $A_m$ and $A_d$, it gives birth to $q$ offspring. The
offspring are located on empty cells in the Moore neighborhood of
the parent by the same way as the SR.

 The age of a newborn baby is set to 1.

(4). Each offspring receives independently its genotype. The way
it is created depends on the kind of population we are
considering. For AR without recombination the baby inherits the
genotype of its parent, change only by mutations. In the case of
SR each offspring receives its own genotype constructed through
simple recombination and one gamete coming from each parent. The
process goes as follows. The genotype of a parent is cut at a
random place. Since the genotype is a double string there are four
pieces which are glued across, forming two gametes. Of these one
is chosen randomly. The same is done for the second parent. This
makes the genotype of the offspring. Then the mutation procedure
is applied.

 Mutations are realized in the following way. On each locus of the
  genotype a mutation may take place with probability $M$.
  The mutation is realized by changing a 1 into 0, or a 0 into 1.

 Phenotypes of the offspring are constructed from the genotypes according to the rule given above.

(5). The individual $k's$ fitness $p_k$, i.e., the degree of
agreement of its phenotype, $h_k$ with the "ideal" one, $H$, is
calculated as the number of bit positions up to $A_k$ where
phenotype and ideal phenotype agree, divided by the current age
$A_k$.

The fitness $p_k$ is compared with a threshold $D$, when it is
lower than this limit, the individual dies.

(6). If the age of the individual $k$ is more than 32, the
individual dies.

(7). After making as many picks as there are in the population at
that time, one Monte Carlo Step has been completed. The age of
every individual is increased by one.

 The changes of the environment are realized as modifications
 of the "ideal" phenotype, i.e. by substituting a 1 by a 0
 in a randomly chosen site (locus), or a 0 by 1, every $T$ MCS.

The parameters of the model are the following: Linear size of the
lattice $L$, the concentration of the initial population
$c(0)=N(0)/( L \times L)$, the minimum age $A_m$ and the maximum
age $A_d$ for reproduction, the maximum number $q$ of offspring
(birth rate), the probability of a mutation per gene for the
babies $M$, the minimum of fitness $D$, the period of changes in
the environment $T$.

Since it would be difficult to operate with so many parameters we
decided on giving to some of them constant values. We took $A_m=8,
A_d=12, D=0.8, c(0)=0.8$. Simulations were performed on a square
lattice of size $L \times L =500 \times 500$. The remaining three
parameters - the birth rate $q$, the mutation probability $M$ and
the period of changes in the environment $T$, have a more
qualitative character.

\section{Results and discussion}

\noindent Let us begin by considering the simplest case-the
environment constant in time. The changes in the environment are
realized here as changes in the "ideal" phenotype with a period
$T$, so the $T$ can be taken as an infinite number here. As shown
in Fig. 1, with parameters $q=2, M=0.01$, both populations adjust
to the environment and reach a high level of concentration. Each
population passes through a deep minimum, which also appears in
Figs. 2, 3 and 4. This is a crucial moment where the population
must overcome the danger of extinction because of stochastic
fluctuations. Recovery is faster for AR, which exhibit its
advantages of efficient and straightforward reproduction (need no
partner to mate) under this condition. The concentration of the
equilibrated population is 0.88 in the asexual and only 0.81 in
the sexual case. The results are similar to those obtained earlier
in the Penna model.

\begin{figure}[htbp]
\vspace*{10pt}\scalebox{0.5}{\includegraphics{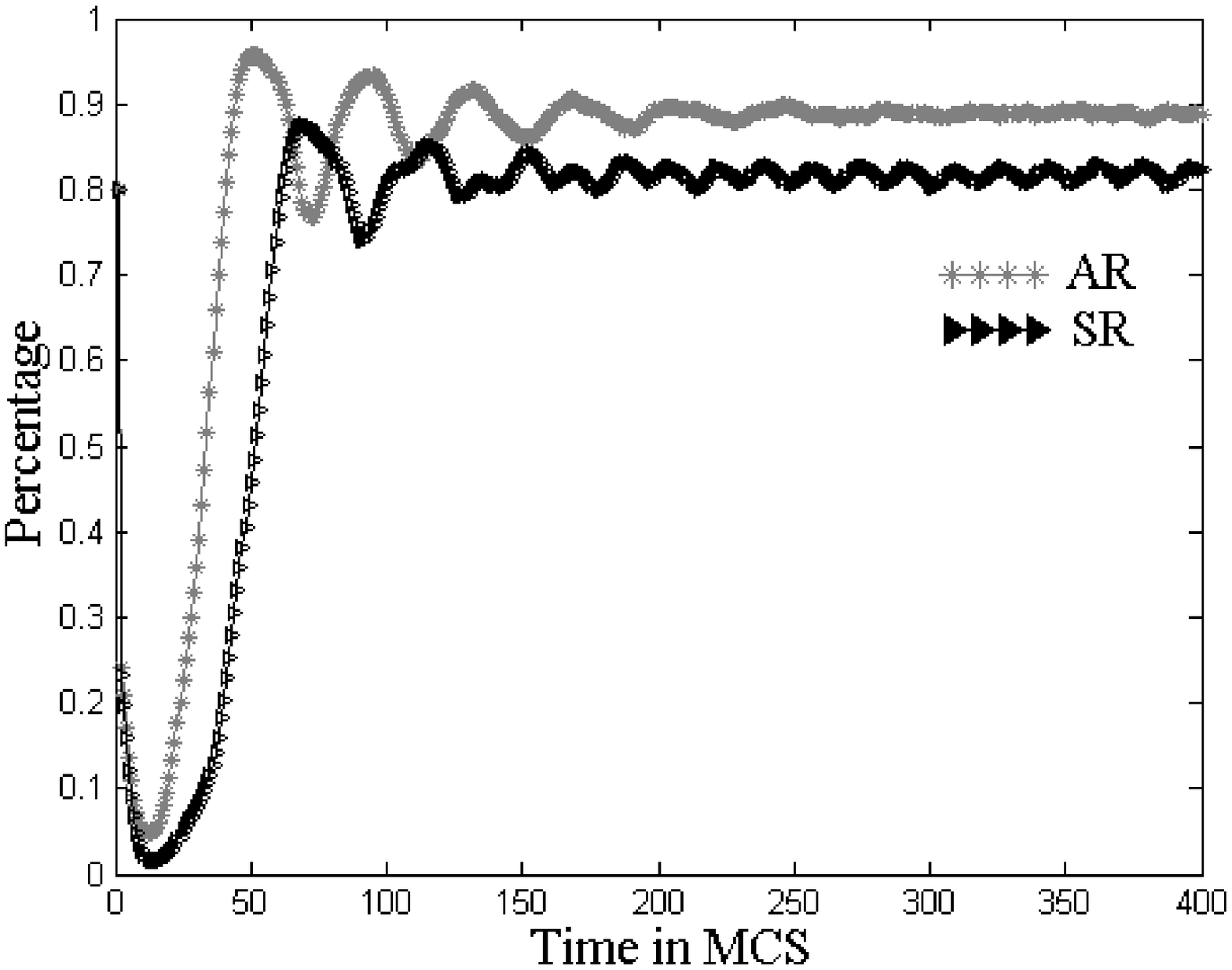}}
 \centering
\fcaption{\footnotesize  Concentration versus time (in MCS). $
q=2, M=0.01.$ } \centerline{\footnotesize Two reproduction
patterns: SR (black triangle) and AR (grey star).}
\end{figure}

When the period of environmental changes is comparable with the
average age of the generation, the effect of environment changes
is pronounced. For $T$=32MCS, using $q=2$, and $M=0.01$, we have
tested that the AR and SR populations may become extinct
frequently, see Fig. 2. With very low birth rate and mutation
probability, both populations cannot adapt well to the changing
environment.

\begin{figure}[htbp]
\vspace*{10pt}\scalebox{0.5}{\includegraphics{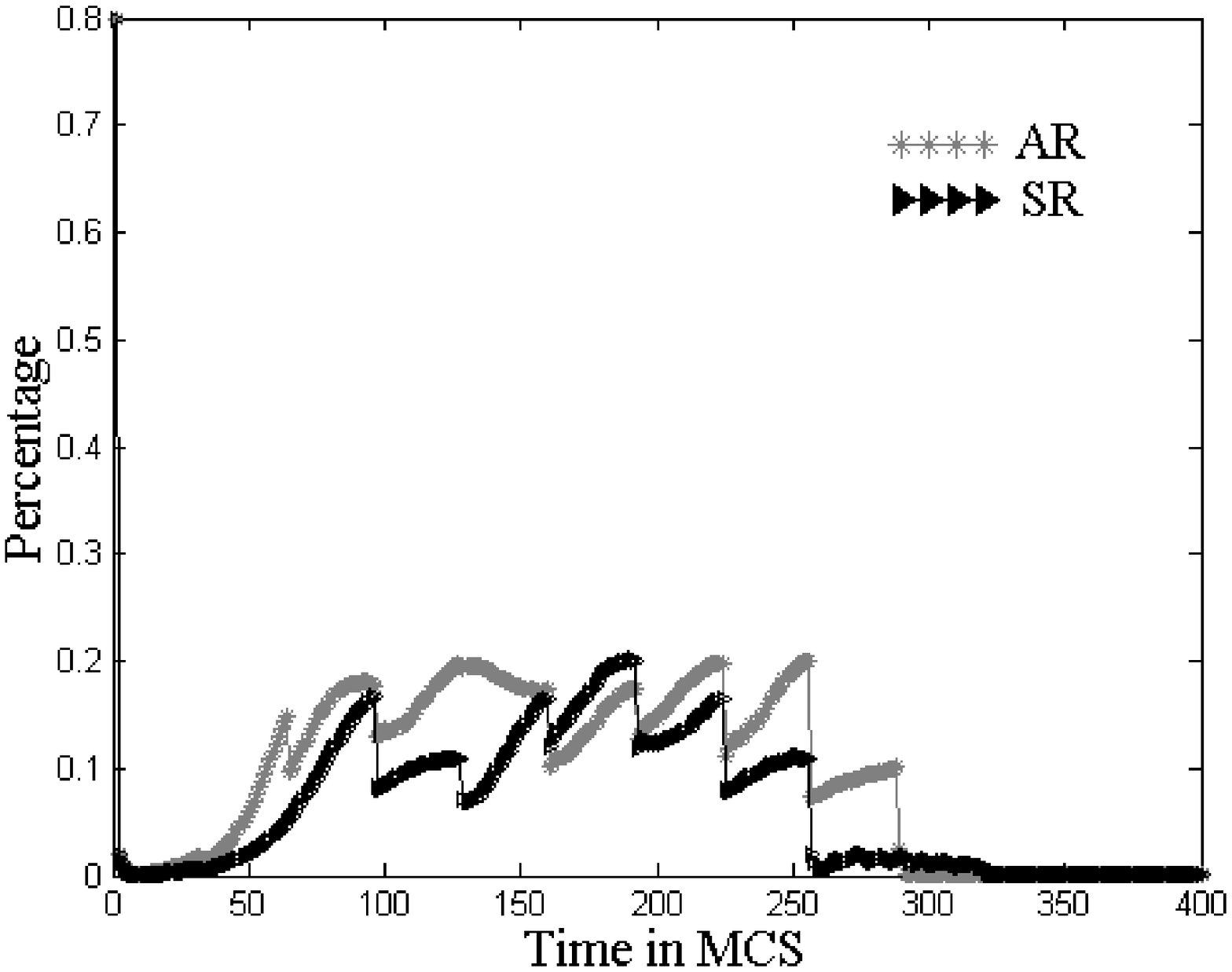}}
 \centering
\fcaption{\footnotesize  Concentration versus time (in MCS).$
T=32, q=2, M=0.01. $ } \centerline{\footnotesize Two reproduction
patterns: SR (black triangle) and AR (grey star).}
\end{figure}

The concentrations of the two populations versus time is shown in
Fig. 3(a) for $T$=8MCS, and in Fig. 3(b) for $T$=32MCS, both
keeping $q=2$ and increasing $M$ from 0.01 to 0.15. For $T$=8MCS
(Fig. 3(a)), although the environment is subject to rapid changes,
the population of AR adapts to the environmental changes and its
concentration fluctuates around a level of 0.41. It should be
however mentioned that the increase of the mutation probability
does not help the SR from extinction. When $T$ is increased to
32MCS (Fig. 3(b)), after surviving the initial sudden drop,
concentrations of the two populations (AR and SR) are mounting
with time towards higher values. For SR, it fluctuates at a
concentration below that of AR. One may conclude that when the
birth rate remains at a low level, increasing $M$ is clearly more
favorable to the development of the population for AR.

\begin{figure}[htbp]
\vspace*{10pt}\scalebox{0.46}{\includegraphics{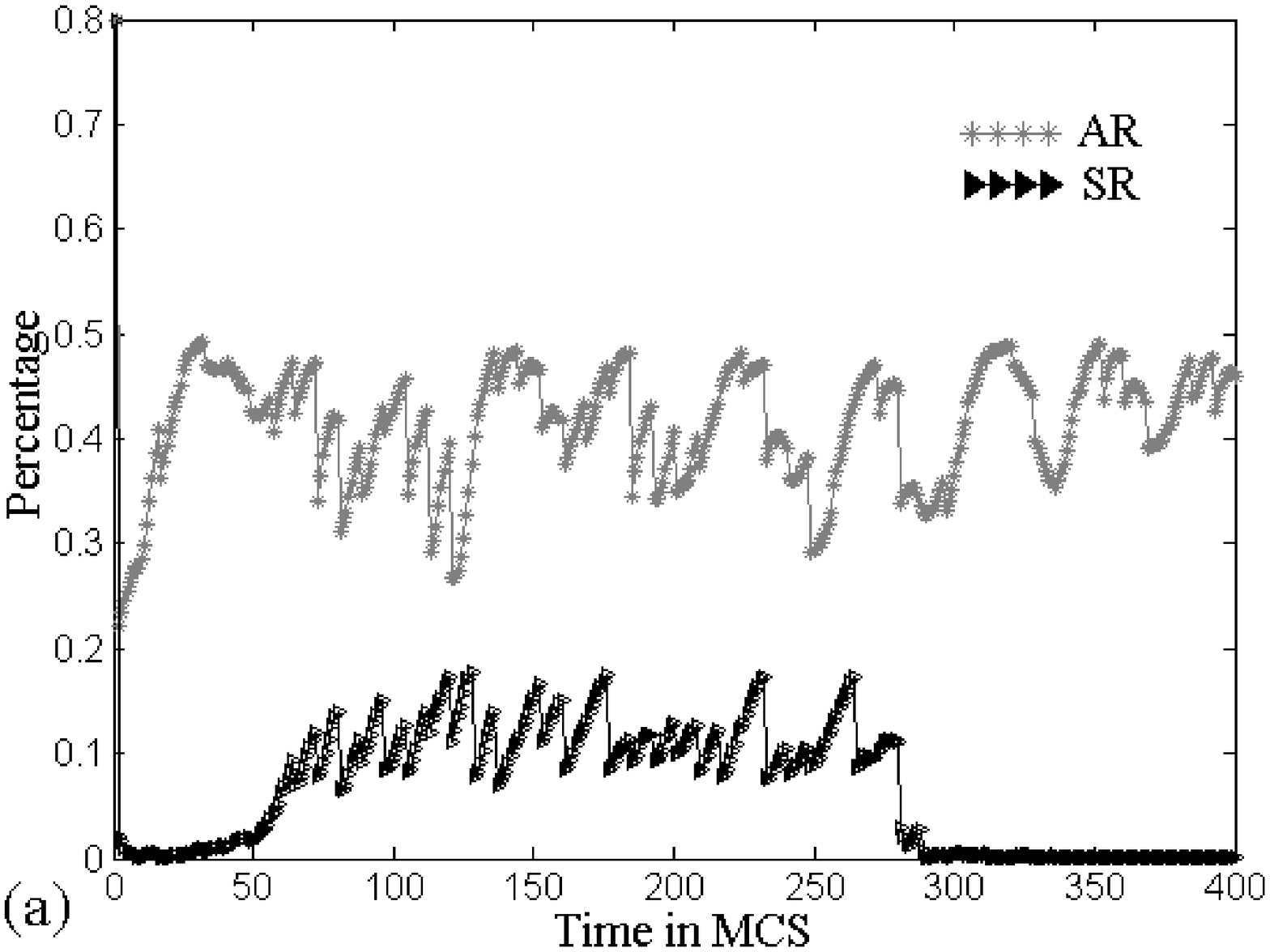}}
 \centering
\vspace*{10pt}\scalebox{0.5}{\includegraphics{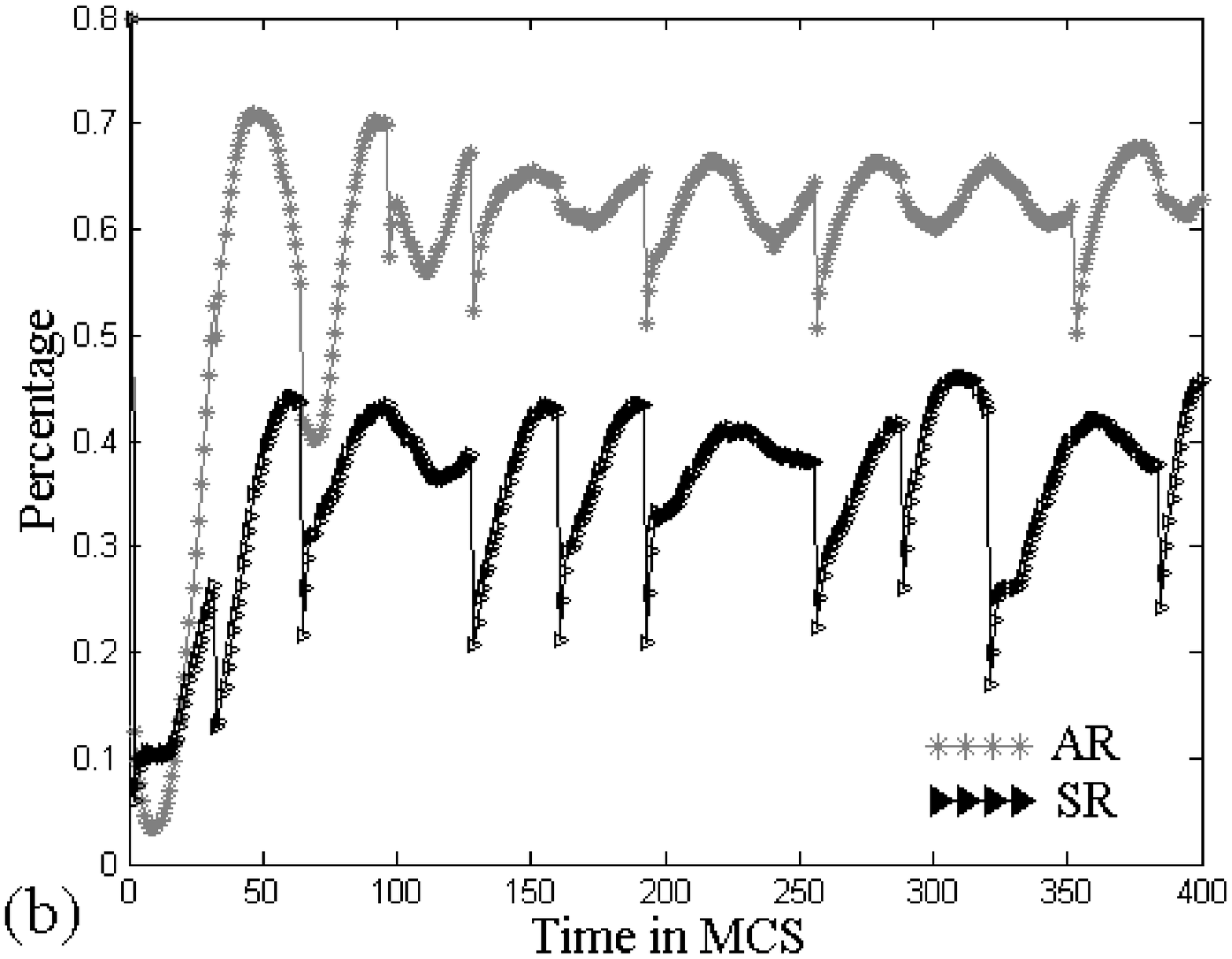}}
 \centering
\fcaption{\footnotesize  Concentration versus time (in MCS).}

 \centerline{\footnotesize (a) $q=2,
M=0.15, T=8.$ (b) $q=2, M=0.15, T=32.$.}

\centerline{\footnotesize Two reproduction patterns: SR (black
triangle) and AR (grey star).}
\end{figure}

The situation changes drastically, in favor of SR, if the birth
rate becomes higher, while the mutation probability remains low.
The effect is presented in Figs. 4(a) and (b). Increase $q$ from 2
to 7, and keep $M=0.01$, then the situation is reversed comparing
to Figs. 3(a) and (b) respectively: the SR is doing better than AR
for both $T$=8 and 32. The benefits of SR are enhanced by higher
birth rate. It can be interpreted as that the effect of
recombination becomes pronounced. But in order to complete
recombination, the individual of SR have to find a partner to
mate, which is demanding. Thus, only the high birth rate, which
may result in a large population, can provide the SR with a
greater chance to mate. In the previous models, the partners for
mating were selected randomly from the whole population. However,
the difficulty of mating for SR, i.e. the spatial limitation, was
not considered.

\begin{figure}[htbp]
\vspace*{10pt}\scalebox{0.46}{\includegraphics{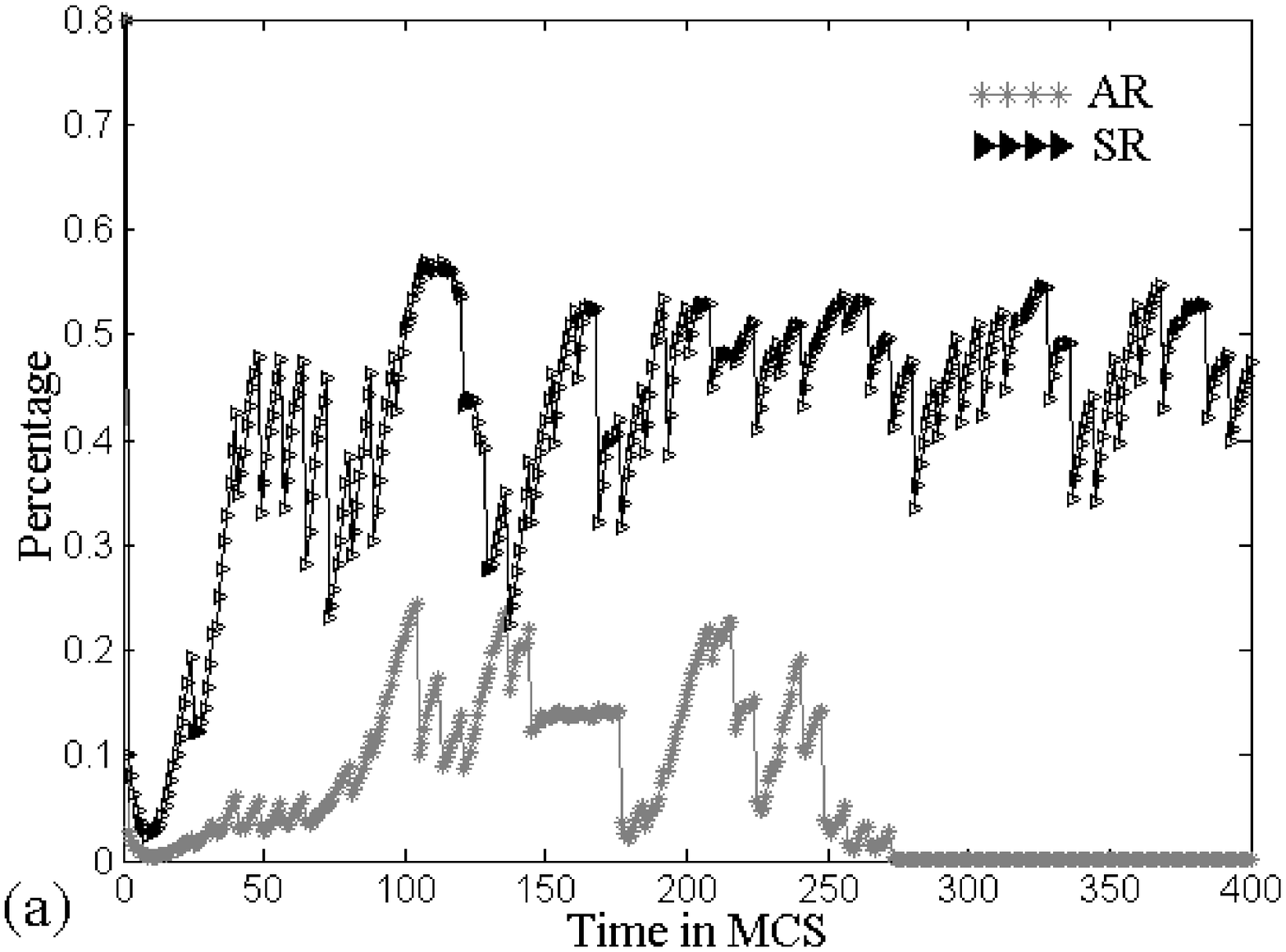}}
 \centering
\vspace*{10pt}\scalebox{0.5}{\includegraphics{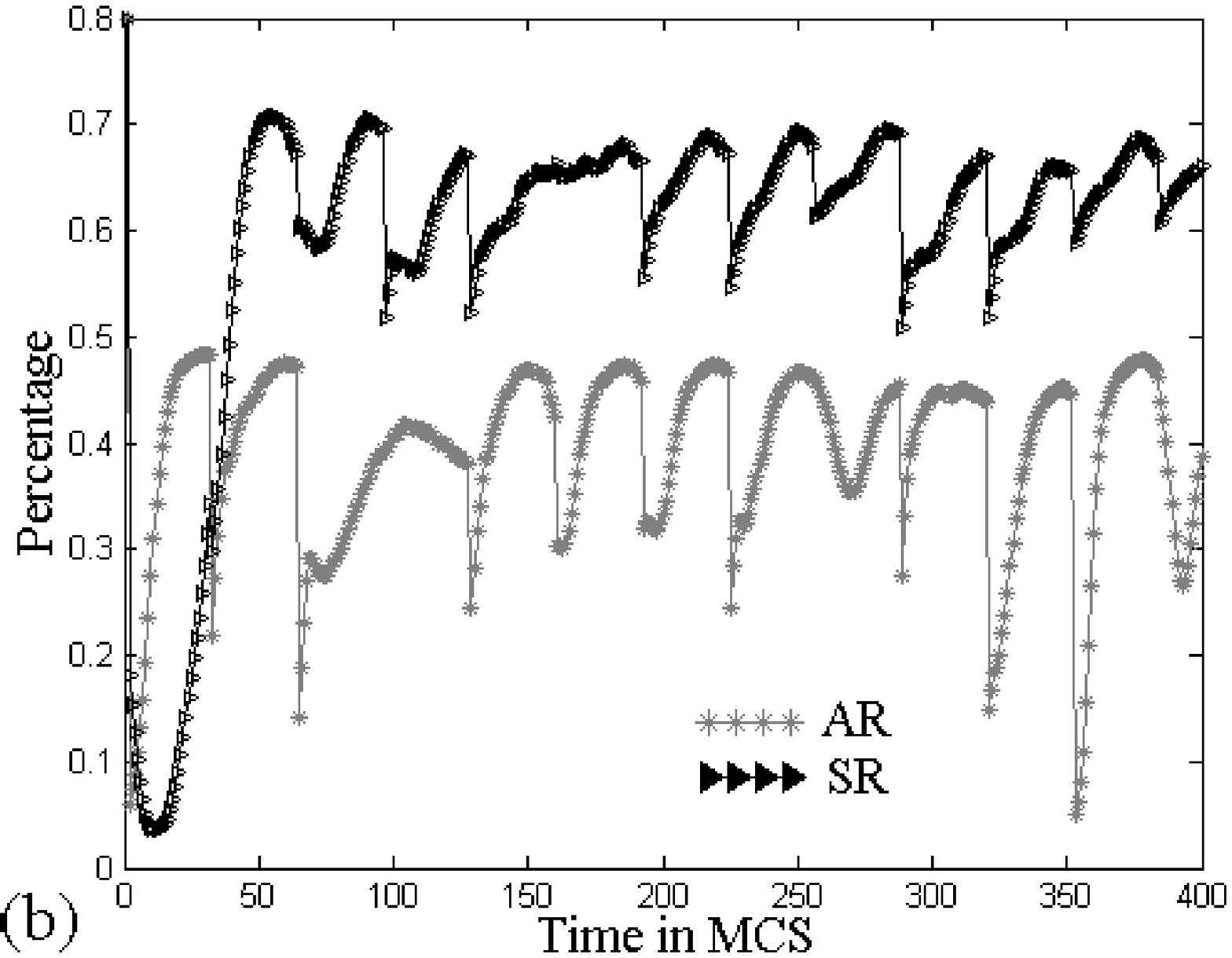}}
 \centering
\fcaption{\footnotesize  Concentration versus time (in MCS).}

 \centerline{\footnotesize (a) $q=7, M=0.01, T=8.$ (b) $q=7, M=0.01, T=32.$.}

\centerline{\footnotesize Two reproduction patterns: SR (black
triangle) and AR (grey star).}
\end{figure}

 As could be expected, when both the birth rate and the mutation
 probability are high, for $T$=8MCS and 32MCS, the populations for
  AR and SR adjust easily to the changes of the environment and
  there is a little preference for AR (the figures are not shown here).

For the evolution of a population the vital question is its
chances of survival in a given environment. Table 1 shows the
percentage of surviving populations after 1000 MCS in a series of
100 runs. As seen from the Table 1, populations with higher
mutation probability and birth rate become extinct less
frequently. For populations with high mutation probability and low
birth rate the AR is the more efficient. When the mutation
probability is low and the birth rate is high, the SR is the
preferred one.

\begin{table}[htbp]
\tcaption{Percentage of surviving populations after 1000 MCS in a
series of 100 runs}

\centerline{\footnotesize\smalllineskip
\begin{tabular}{c c c c  c}\\
\hline
{}  &{T=8} &T=8 &T=32 &T=32\\
\hline
 &AR &SR &AR &SR\\
\hline
q=2, M=0.01 &0.05 &0.04 &0.08 &0.09 \\
q=2, M=0.15 &0.71 &0.45 &0.82 &0.56  \\
q=7, M=0.01 &0.48 &0.80&0.54 &0.84\\
q=7, M=0.15 &0.97 &0.93 &0.99 &0.94 \\
\hline\\
\end{tabular}}
\end{table}

Thus, depending on parameters, either AR or SR can be the clearly
preferred choice, in agreement with reality. It is well known that
the advantages of AR are efficient and straightforward, here we do
not discuss it further and concentrate on the SR.

\begin{figure}[htbp]
\vspace*{10pt}\scalebox{0.46}{\includegraphics{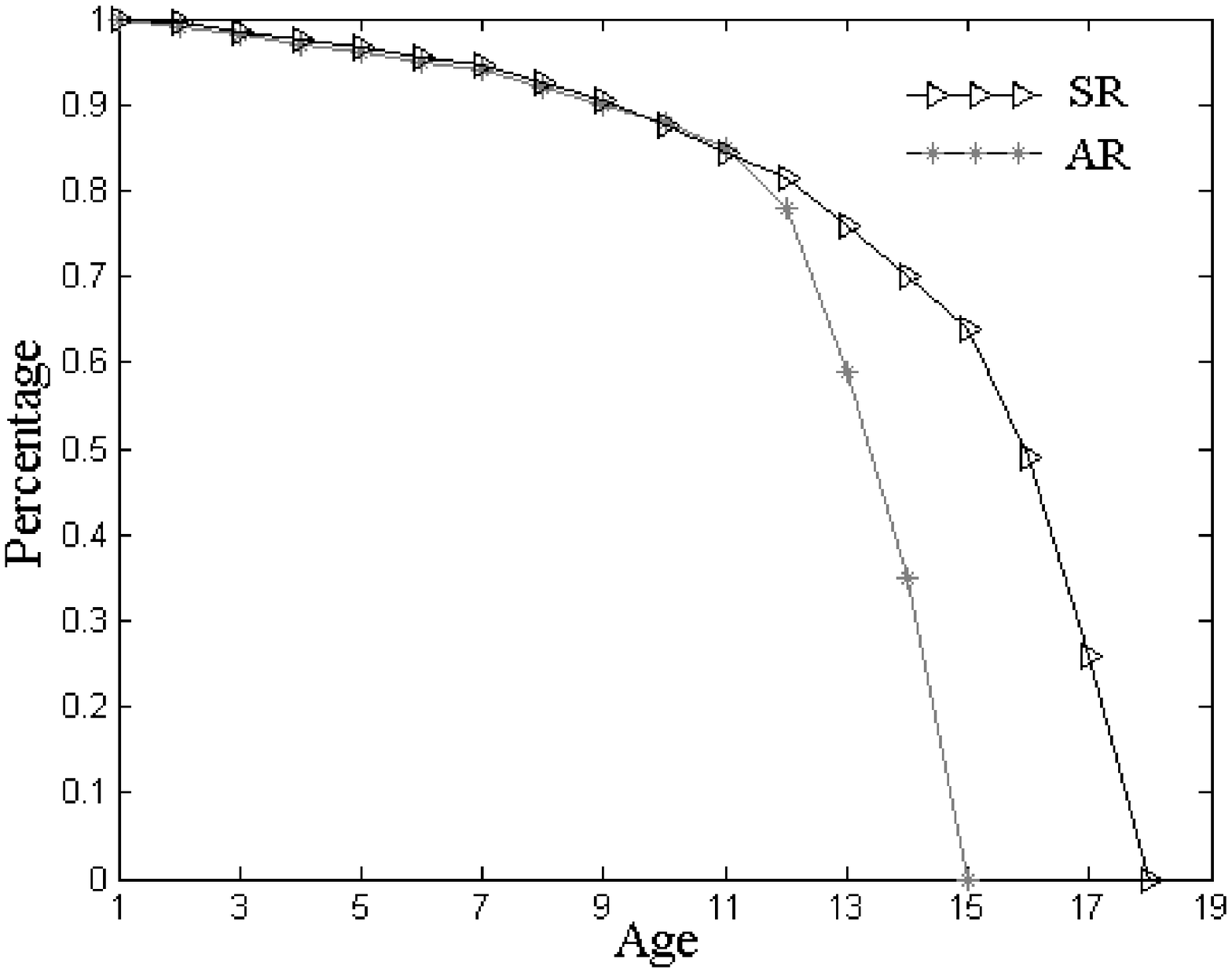}}
 \centering

\fcaption{\footnotesize  Age distribution (in percentage) of the
AR and SR populations }

 \centerline{\footnotesize versus age dependence. Parameters:$ q=7, M=0.01$ and $T$=8.}

\centerline{\footnotesize Two reproduction patterns: SR (black
triangle) and AR (grey star).}
\end{figure}

For $q=7, M=0.01$ and $T=8$, we present the age structure of
populations for AR and SR in the steady state, see Fig. 5. The
y-coordinate is the ratio between the number of the individuals
with a certain age and the number of the whole population. The
averaging was done over 50 years.

In case of the AR we can see that the curve goes down slowly, and
then, after the maximum reproductive age there is a sudden drop,
which has been also mentioned in $^{13,14}$. This phenomenon can
be explained as follows. Although the environment is changing,
most of the genes on the phenotype which do not adapt to the
environment are able to accumulate beyond the maximum reproduction
age in time. As to the SR case, the drop is delayed about 3 years
and the drop is less steep. Sexual reproduction also ensures that
the slowly decreased stage before the drop is maintained longer.

Sexually reproducing individuals, males and females, live longer
than the asexually reproducing females. Why can the females
survive after their reproductive age is passed in the case of SR
while they cannot in the AR case? Since the sex is not transmitted
genetically, we take each child as male or female with probability
0.5. Then either both males and females die soon, or both males
and females die late. Thus, males help the survival of younger
females beyond their reproduction age.

The above conclusion has some resemblance to the ones obtained in
$^{9}$. The difference is that here the curve decreases before the
sudden drop because of the pressure of environmental changes,
while in $^{9}$ there was hardly any decrease.

\section{Conclusions}
\noindent The present paper first introduces the following
assumptions. In the constantly changing environment, we assume
that the mutations are blind. The mutations which are helpful to
make the population more adaptive to the environment are called
beneficial mutations, and those which are negative for the
population to adjust to the environment are called deleterious
mutations.

Basing on the above assumptions, we presented a lattice model of
population dynamics based on Monte Carlo simulations. Each
individual is characterized, apart from its location on the
lattice, by its genotype, phenotype and age. The agreement of the
phenotype with the "ideal" phenotype characterizing the
environment determines the individual fitness. There are three
major factors influencing the evolving process-birth rate,
mutation probability and the period of changes in the environment.
Depending on the three parameters, either AR or SR can be
favorable.

With very low birth rate and mutation probability, both
populations cannot adapt well to the changing environment. The AR
and SR populations may become extinct frequently. If the mutation
probability is high and the birth rate remains at a low level, the
advantage of AR on the adaptation to the changing environment is
more pronounced, which is ascribed to its efficient and
straightforward nature (need no partner to mate) reproduction
scheme. And this preference of the AR disappears if the birth rate
of the population is high and the mutation probability remains at
a low level. Then the SR is the more favorable, in agreement with
the conclusion of $^{15}$. The present work supports the claims
that SR by keeping more diversified populations gives them better
chance to survive cataclysms. However, in order to obtain a
greater chance to mate, the birth rate should be high.

In our simulations the populations with different reproduction
schemes were not evolving together, hence there was no effect of
direct competition. However both evolution processes started from
the same conditions, like initial concentration of individuals,
distribution of the genotypes and habitat requirements.

\nonumsection{References}


\noindent

\begin{thebibliography}{000}
\bibitem{1}
A.S. Kondrashov, {\bibit Nature} {\bibbf  369}, 99 (1994).

\bibitem{2}
H.J. Muller, Am., {\bibit Nature} {\bibbf  66}, 118 (1932).

\bibitem{3}
B. Orcal, E. Tuzel, V. Sevim, N. Jan, A. Erzan, {\bibit Int. J.
Mod. Phys. C} {\bibbf  11}, 9736 (2000).

\bibitem{4}
E. Tuzel, V. Sevim, A.Erzan, eprint cond-mat/0101426.

\bibitem{5}
S. Moss de Oliveira, P.M.C. de Oliveira, D. Stauffer, {\bibit
Evolution, Money, War and Computers },(Teubner, Leipzig, 1999).

\bibitem{6}
 T.J.P. Penna, {\bibit J. Stat. Phys.} {\bibbf  78},1629 (1995).


\bibitem{7}
J.S. Sa Martins, S. Moss de Oliveira, {\bibit Int. J. Mod. Phys.
C} {\bibbf  9}, 421 (1998).

\bibitem{8}
K.W. Wachter, C.E. Finch, Between Zeus and the Salmon,{\bibit The
Biodemography of Longevity} ,  (National Academy Press,
Washington, DC, 1997).

\bibitem{9}
D. Stauffer, P.M.C. de Oliveira, S. Moss de Oliveira, T.J.P. Penna
and J.S. Sa Martins, arXiv:cond-mat/0011524 v1 (2000)

\bibitem{10} I. Mroz, A. Pekalski, K. Sznajd-Weron,  {\bibit
Phys. Rev. Lett.} {\bibbf  76}, 3025 (1996).

\bibitem{11}
A. Fraser, D. Burnell,{\bibit Computer Models in Genetics},
(McGraw-Hill,  New York, 1970).

\bibitem{12}
D. Charlesworth, M.T. Morgan, B. Charlesworth, {\bibit Genet. Res.
} (Cambridge) 59, 49 (1992).

\bibitem{13}
T.J.P. Penna, S. Moss de Oliveira and D. Stauffer, {\bibit
Phys.Rev. E } {\bibbf  52}, 3309 (1995).

\bibitem{14}
 T.J.P. Penna and S. Moss de Oliveira,{\bibit J. Physique I } {\bibbf  5},
1697 (1995).

\bibitem{15}
A. Pekalski, {\bibit Eur. Phys. J. B} {\bibbf  13}, 791 (2000).




\end{thebibliography}
\end{document}